\documentclass[traditabstract,longauth]{aa}  

\usepackage{natbib}
\bibpunct{(}{)}{;}{a}{}{,} % to follow the A&A style
\usepackage{graphicx}
\usepackage{txfonts}
\newcommand{\upd}[1]{#1}
\newcommand{\updB}[1]{#1}

\begin{document} 

   \title{\emph{Herschel} detects oxygen in the $\beta$~Pictoris debris disk
   \thanks{\emph{Herschel} is an ESA space observatory with science
     instruments provided by European-led Principal Investigator
     consortia and with important participation from NASA.} }

\author{
A.\,Brandeker\inst{\ref{inst4}},
G.\,Cataldi\inst{\ref{inst4}},
G.\,Olofsson\inst{\ref{inst4}},
B.\,Vandenbussche\inst{\ref{inst1}},
B.\,Acke\inst{\ref{inst1}},
M.\,J.\,Barlow\inst{\ref{inst7}},
J.\,A.\,D.\,L.\,Blommaert\inst{\ref{inst1},\ref{inst24}},
M.\,Cohen\inst{\ref{inst9}},
W.\,R.\,F.\,Dent\inst{\ref{inst10}},
C.\,Dominik\inst{\ref{inst5},\ref{inst6}},
J.\,Di\,Francesco\inst{\ref{inst11}},
M.\,Fridlund\inst{\ref{inst21}},
W.\,K.\,Gear\inst{\ref{inst13}},
A.\,M.\,Glauser\inst{\ref{inst14},\ref{inst2}},
J.\,S.\,Greaves\inst{\ref{inst15}},
P.\,M.\,Harvey\inst{\ref{inst16}},
A.\,M.\,Heras\inst{\ref{inst12}},
M.\,R.\,Hogerheijde\inst{\ref{inst18}},
W.\,S.\,Holland\inst{\ref{inst2}},
R.\,Huygen\inst{\ref{inst1}},
R.\,J.\,Ivison\inst{\ref{inst23},\ref{inst19}},
S.\,J.\,Leeks\inst{\ref{inst20}},
T.\,L.\,Lim\inst{\ref{inst20}},
R.\,Liseau\inst{\ref{inst21}},
B.\,C.\,Matthews\inst{\ref{inst11}},
E.\,Pantin\inst{\ref{inst3}},
G.\,L.\,Pilbratt\inst{\ref{inst12}},
P.\,Royer\inst{\ref{inst1}},
B.\,Sibthorpe\inst{\ref{inst22}},
C.\,Waelkens\inst{\ref{inst1}},
H.\,J.\,Walker\inst{\ref{inst20}}
}

\institute{
Department\ of\ Astronomy,\ Stockholm\ University,\ AlbaNova\ University\ Center,\ 106\,91\ Stockholm,\ Sweden\\
\email{alexis@astro.su.se}\label{inst4}
\and % 5 
Instituut\ voor\ Sterrenkunde,\ KU\ Leuven,\ Celestijnenlaan\ 200\ D,\ B-3001\ Leuven,\ Belgium\label{inst1}
\and % 7
Department\ of\ Physics\ and\ Astronomy,\ University\ College\ London,\ Gower\ St,\ London\ WC1E\ 6BT,\ UK\label{inst7}
\and % 7
Astronomy\ and\ Astrophysics\ Research\ Group,\ Dep.\ of\ Physics\ and\ Astrophysics,\ Vrije\ Universiteit\ Brussel,\ Pleinlaan\ 2,\ 1050\ Brussels,\ Belgium\label{inst24}
\and % 9
Radio\ Astronomy\ Laboratory,\ University\ of\ California\ at\ Berkeley,\ CA\ 94720,\ USA\label{inst9}
\and % 10 
ALMA,\ Alonso\ de\ C\'ordova\ 3107,\ Vitacura,\ Santiago,\ Chile\label{inst10}
\and % 4
Astronomical\ Institute\ Anton\ Pannekoek,\ University\ of\ Amsterdam,\ Kruislaan\ 403,\ 1098\ SJ\ Amsterdam,\ The\ Netherlands\label{inst5}
\and % 6 
Afdeling\ Sterrenkunde,\ Radboud\ Universiteit\ Nijmegen,\ Postbus\ 9010,\ 6500\ GL\ Nijmegen,\ The\ Netherlands\label{inst6}
\and % 11 
National\ Research\ Council\ of\ Canada,\ Herzberg\ Institute\ of\ Astrophysics,\ 5071\ West\ Saanich\ Road,\ Victoria,\ BC,\ V9E\ 2E7,\ Canada\label{inst11}
\and % 21
Earth\ and\ Space\ Sciences,\ Chalmers\ University\ of\ Technology,\ SE-412\ 96\ Gothenburg,\ Sweden\label{inst21}
\and % 13
School\ of\ Physics\ and\ Astronomy,\ Cardiff\ University,\ Queens\ Buildings\ The\ Parade,\ Cardiff\ CF24\ 3AA,\ UK\label{inst13}
\and % 14
Institute\ of\ Astronomy,\ ETH\ Zurich,\ 8093\ Zurich,\ Switzerland\label{inst14}
\and % 2
UK Astronomy Technology Centre, Royal Observatory Edinburgh, Blackford Hill, EH9 3HJ, UK\label{inst2}
\and % 15 
School\ of\ Physics\ and\ Astronomy,\ University\ of\ St\ Andrews,\ North\ Haugh,\ St\ Andrews,\ Fife\ KY16\ 9SS,\ UK\label{inst15}
\and % 16
Department\ of\ Astronomy,\ University\ of\ Texas,\ 1\ University\ Station\ C1400,\ Austin,\ TX 78712,\ USA\label{inst16}
%\and % 17
%CASA,\,University\,of\,Colorado,\,389-UCB,\,Boulder,\,CO 80309,\,USA\label{inst17}
\and % 12
ESA,\ Directorate\ of\ Science,\ Scientific\ Support\ Office,\ European\ Space\ Research\ and\ Technology\ Centre\ (ESTEC/SCI-S),\ Keplerlaan\ 1,\ NL-2201\ AZ\ Noordwijk,\ The\ Netherlands\label{inst12}
\and % 18
Leiden\ Observatory,\ Leiden\ University,\ PO\ Box\ 9513,\ 2300\ RA,\ Leiden,\ The Netherlands\label{inst18}
\and % 23
European\ Southern\ Observatory,\ Karl-Schwarzschild-Strasse\ 2,\ 85748\ Garching,\ Germany\label{inst23}
\and % 19
Institute\ for\ Astronomy,\ University\ of\ Edinburgh,\ Blackford\ Hill,\ Edinburgh\ EH9\ 3HJ,\ UK\label{inst19}
\and % 20
Space\ Science\ and\ Technology\ Department,\ Rutherford\ Appleton\ Laboratory,\ Oxfordshire,\ OX11\ 0QX,\ UK\label{inst20}
\and % 3
Laboratoire\ AIM,\ CEA/DSM-CNRS-Universit\'e\ Paris\ Diderot,\ IRFU/Service\ d'Astrophysique,\ Bat.709,\ CEA-Saclay,\ 91191\ Gifsur-Yvette\ Cedex,\ France\label{inst3}
\and 
SRON\ Netherlands\ Institute\ for\ Space\ Research,\ P.O.\ Box\ 800,\ 9700\ AV\ Groningen,\ The\ Netherlands\label{inst22}
%\and % 22
%Institute for Space Imaging Science,\ University\ of\ Lethbridge,\ Lethbridge,\ Alberta,\ T1J\ 1B1,\ Canada\label{inst22}
%\and % 8
%Max-Planck-Institut\ f\"ur Astronomie,\ K\"onigstuhl\ 17,\ D-69117\ Heidelberg,\ Germany\label{inst8}
}

 \date{Received ; accepted}

% \abstract{}{}{}{}{} 
% 5 {} token are mandatory
 
  \abstract{The young star $\beta$~Pictoris is well known for its dusty debris disk, produced through the grinding down by collisions of planetesimals, kilometre-sized bodies in orbit around the star. In addition to dust, small amounts of gas are also known to orbit the star, likely the result from vaporisation of violently colliding dust grains. The disk is seen edge on and from previous absorption spectroscopy we know that the gas is very rich in carbon relative to other elements. The oxygen content has been more difficult to assess, however, with early estimates finding very little oxygen in the gas at a C/O ratio $20\times$ higher than the cosmic value. A C/O ratio that high is difficult to explain and would have far-reaching consequences for planet formation. Here we report on observations by the far-infrared space telescope \textit{Herschel}, using PACS, of emission lines from ionised carbon and neutral oxygen. The detected emission from C$^+$ is consistent with that previously reported being observed by the HIFI instrument on \textit{Herschel}, while the emission from O is hard to explain without assuming a higher-density region in the disk, perhaps in the shape of a clump or a dense torus, required to sufficiently excite the O atoms. A possible scenario is that the C/O gas is produced by the same process responsible for the CO clump recently observed by ALMA in the disk, and that the re-distribution of the gas takes longer than previously assumed. A more detailed estimate of the C/O ratio and the mass of O will have to await better constraints on the C/O gas spatial distribution.
}
 

\defcitealias{rob06}{ROB06}
\defcitealias{bra04}{BLOM04}
\defcitealias{zag10}{ZBW10}

\authorrunning{A.~Brandeker et al.}
\titlerunning{\emph{Herschel} detects O in the $\beta$\,Pic disk}

   \maketitle

%
%________________________________________________________________

\section{Introduction}

The star $\beta$~Pictoris is young \citep[23$\pm$3\,Myr,][]{mam14}, nearby \citep[19\,pc,][]{van07}, and harbours a large debris disk, making it a target of intense scrutiny since its discovery by the infrared astronomical satellite \textit{IRAS} in 1984 \citep{aum85}. As the survival time for dust grains in the disk is far shorter than the age of the system, it was argued early on that the dust must be replenished through collisional fragmentation of larger bodies -- hence the name `debris disk' \citep{bac93}. The interest in the $\beta$\,Pic disk is strongly linked to the interest in planet formation -- the disk was discovered a decade before the first exoplanets, yet is apparently the result of the same mechanisms that form planets, and so provides a valuable example to test theory. Rocky planets, like the Earth, are generally believed to be built up by smaller bodies called planetesimals, which range in size from 1 kilometre to hundreds of kilometres \citep{nag07}. How the planetesimals are built up, in turn, is one of the outstanding problems of planet formation theory today, with one of the more promising suggestions being production through a `streaming instability' \citep{joh07}. 

Why some stars seem to form planets while others may not is still not understood, but one clue could be the observed correlation between planet incidence and the elemental abundances of the parent star: stars of higher heavy element abundance are argued to be more likely to have massive planets \citep{fis05,buc14}. If the composition of the star reflects the composition of the planet-forming disk, one explanation could be that the dust grains that eventually build up the planetesimals are more easily formed in an environment enriched in heavier elements. The picture is far from clear, however, since a correlation with host star metallicity is observed for gas giants, but not for small exoplanets \citep{buc14} or the tracers of planetesimal formation, debris disks \citep{gre06,wya07,mor15}.

The $\beta$\,Pic system has passed its \upd{planetesimal formation phase and today we observe a resulting} planetesimal belt located at a radial distance of 100\,AU \citep{the07,wil11,den14}, and a recently discovered massive planet (10--12 times Jupiter's mass) on a $\sim$9\,AU orbit \citep{lag10,mil15}. In addition to the dust produced through a collisional cascade originating from the planetesimals, we also observe gas in the system \citep[hereafter \citetalias{rob06}]{hob85,olo01,rob06}. With an estimated total mass just a fraction of an Earth mass \citep[][hereafter \citetalias{zag10}]{zag10}, this gas is far too tenuous to contribute to the formation of new planets. Instead, it is thought to be the result of the reverse process, vaporisation of colliding dust grains originating in the planetesimals \citep{lis98,cze07}. Alternatively, the gas could also be released from the dust through photo-desorption \citep{che07,gri07} or collisions between volatile-rich comets in a massive Kuiper-belt analogue \citep{zuc12}. 
The latter seems favoured by recent
 ALMA observations of CO apparently released from a clump located at 85\,AU, where an enhanced collision rate induced by either a resonance trap from a migrating planet or a residue from a former giant collision are suggested as possible mechanisms \citep{den14,jac14}. Given the rapid dissociation time for CO in this environment \citep[120\,yr; ][]{vis09}, it is clear that the gas must be currently produced, and that it is a source of both C and O.

The long-standing puzzle of how gas could be kept in the disk while subject to a strong radiation force from the star was resolved by the discovery of a large overabundance of carbon with respect to \updB{detected} metallic elements in the disk gas \citepalias[e.g., Na, Fe, and Ca;][]{rob06}, acting as a braking agent \citep{fer06}.
The overabundance is not necessarily a consequence of the dust grains being unusually carbon rich, it could also be due to chemical differentiation through preferential removal of the observed metallic elements \citep{xie13}. In contrast to carbon, the metallic elements experience a strong radiation force from the star, up to a few hundred times stronger than the gravitational force. Oxygen, on the other hand, is similar to carbon in that it is not affected by radiation pressure from the star \citep{fer06}, and is thus expected to be closely mixed with carbon. It was therefore surprising when absorption spectroscopy found a C/O ratio $20\times$ higher than the cosmic abundance found in the Sun, and $100\times$ the ratio found in carbonaceous chondrite meteorites \citepalias{rob06}. 

The C/O ratio is believed to strongly affect the outcome of the planet formation process \citep{kuc05}. For example, the sequence by which elements condense as the gas cools, and thus the mass distribution and planet formation efficiency, changes significantly with the C/O ratio. A ratio larger than 0.8 would result in carbide-dominated interiors of planets \citep{bon10}, as opposed to the silicate-dominated composition found in the rocky planets of our solar system. An example of an extra-solar planet where a super-cosmic C/O ratio has been suggested is WASP-12b, where already a $2\times$ higher C/O ratio in its atmosphere results in dramatically different mixing ratios of molecular species \citep{mad11}. If the C/O ratio of the disk gas reflects the composition of the
$\beta$\,Pic planet,  and the C/O ratio is indeed higher than the cosmic ratio by a factor 20, then the planet would likely have a diamond core and be evidence for a very exotic planet formation scenario \citep{kuc05}. The evolutionary and atmospheric models used to estimate the mass of the planet from photometry would be invalid \citep{lag10,bar03}, and would have to be replaced with alternative, high-C/O models. To determine the relative abundance of C and O is thus highly relevant for understanding the available paths for planet formation.
We here present observations of \ion{C}{ii} and \ion{O}{i}, obtained using the far-infrared space telescope \textit{Herschel}, that aim to enable an estimate of the O mass and C/O ratio of the gas in the $\beta$\,Pic debris disk.

\begin{table}
\caption{\label{t:flux}PACS detected emission from the $\beta$\,Pic gas disk}
\centering
\begin{tabular}{lcc}
\hline\hline
Line & Spaxel & Flux/beam\tablefootmark{a} \\
 &  & $[$erg\,s$^{-1}$cm$^{-2}$\,beam$^{-1}]$\\
\hline
$[$\ion{O}{i}$]$ 63.2\,$\mu$m & 12\tablefootmark{b} & $(8.65\pm1.70)\times10^{-15}$ \\
$[$\ion{C}{ii}$]$ 157.7\,$\mu$m & 12\tablefootmark{b} & $(1.73\pm0.17)\times10^{-14}$ \\
$[$\ion{C}{ii}$]$ 157.7\,$\mu$m & 11 & $(4.32\pm1.31)\times10^{-15}$ \\
$[$\ion{C}{ii}$]$ 157.7\,$\mu$m & 13 & $(3.91\pm1.41)\times10^{-15}$ \\
$[$\ion{C}{ii}$]$ 157.7\,$\mu$m & 17 & $(6.42\pm1.73)\times10^{-15}$ \\
\hline
\end{tabular}
\tablefoot{
\tablefoottext{a}{Beam sizes are approximately 9.5\arcsec\ and 11\arcsec\ (full width at half maximum) for [\ion{O}{i}] and [\ion{C}{ii}], respectively.
\upd{The quoted errors are random. The absolute flux calibration is expected to be better than 30\,\% (see Sect.\,\ref{s:results}).}}
\tablefoottext{b}{Central spaxel.}
}
\end{table}

\begin{figure}
   \centering
   \includegraphics[width=8.5cm]{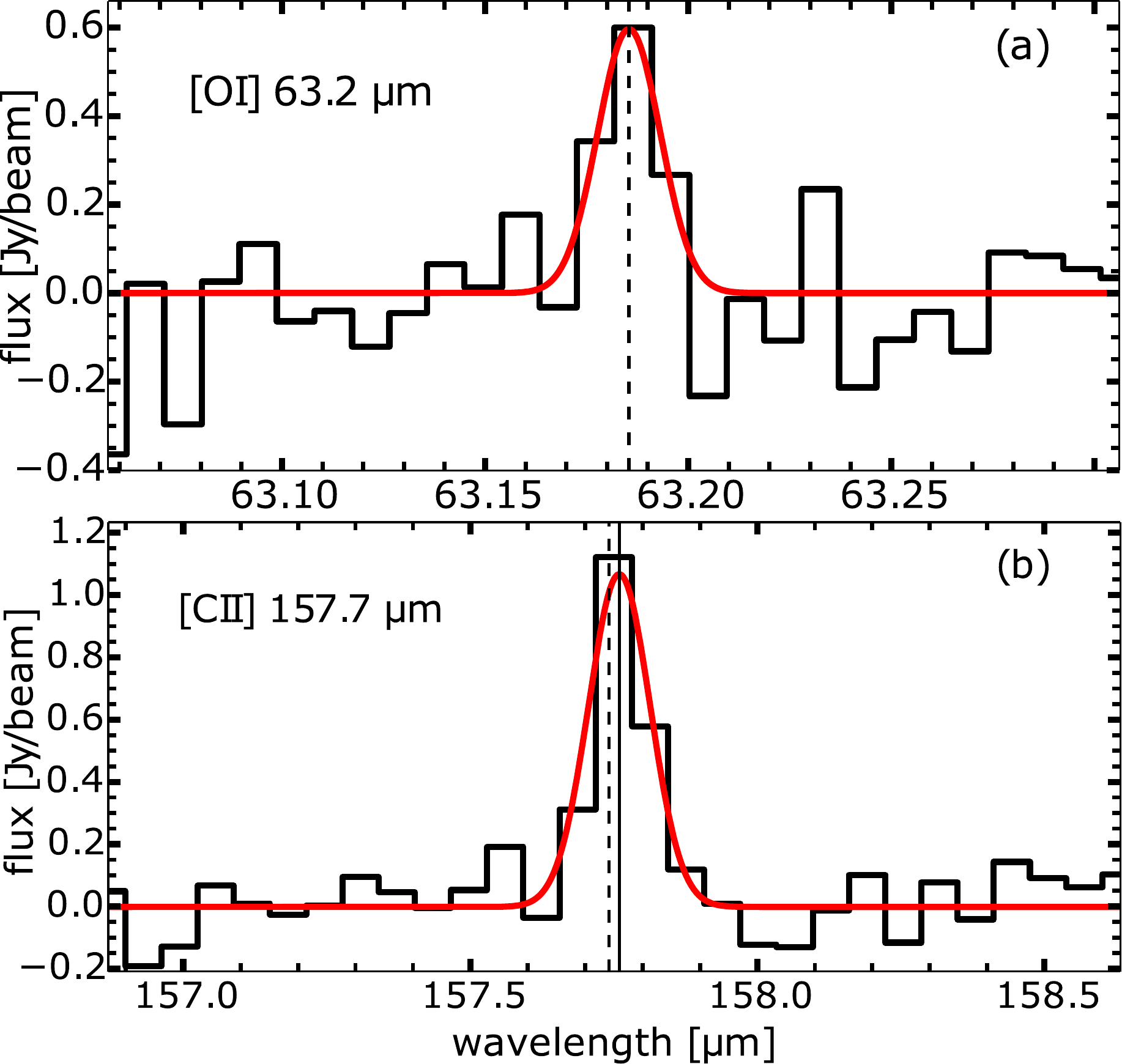}
      \caption{Emission lines observed from $\beta$\,Pic, with fitted Gaussian profiles overplotted. \upd{The spectra have been continuum subtracted and rebinned.} The vertical dashed line
shows the expected wavelength of the emission line.The measured fluxes are reported in Table~\ref{t:flux}. 
Upper panel (a): The [\ion{O}{i}] 63.2\,$\mu$m emission line observed by the central spaxel (12) of PACS. The wavelength scale is in the local standard of rest frame.
Lower panel (b): The [\ion{C}{ii}] 157.7\,$\mu$m emission line of the central spaxel. The unbroken vertical line shows the fitted centre of the emission line.}
         \label{f:ciioi}
   \end{figure}

\begin{figure}
   \centering
   \includegraphics[width=8.5cm]{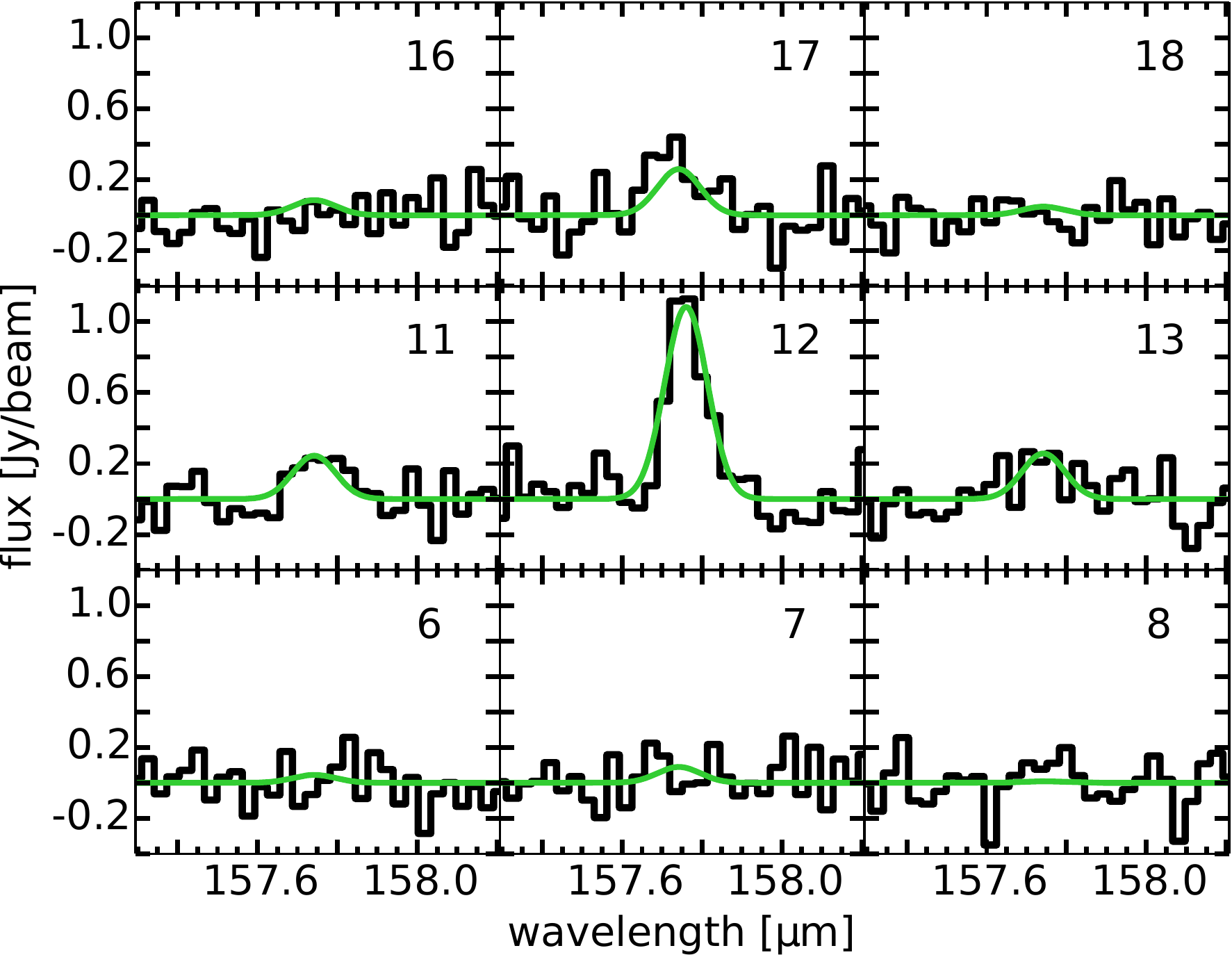}
      \caption{The central 3$\times$3 spaxels of the 5$\times$5 PACS array, showing the detected [\ion{C}{ii}] 157.7\,$\mu$m emission line. Spaxels outside this region show no signal. In addition to the central spaxel 12, the adjacent spaxels 11, 13 and 17 also detect a signal (listed in Table~\ref{t:flux}). Overplotted on the data are synthetic observations of the disk model `$5\times5$\,AU' (Sect.\,\ref{s:model}, Table~\ref{t:model}).}
   \label{f:3x3}
   \end{figure}

\section{Observations}

The data presented are part of the \textit{Herschel} guaranteed time \textit{Stellar Disk Evolution} key programme (PI Olofsson; OBSIDs 1342188425 and 1342198171, observed on 2009-12-22 [\ion{O}{i}] and 2010-06-02 [\ion{C}{ii}], respectively). We used the Photodetector Array Camera and Spectrometer \citep[PACS,][]{pog10} on-board the \textit{Herschel Space Observatory} \citep{pil10}, operating as an integral field spectrometer to observe $\beta$\,Pic in the 158\,$\mu$m [\ion{C}{ii}] and 63\,$\mu$m [\ion{O}{i}] line regions. 
The [\ion{O}{i}] 63\,$\mu$m line was observed with a dedicated PACS chop/nod line spectroscopy observation centred on the line, while the  [\ion{C}{ii}] 158\,$\mu$m line was extracted from a deep observation of the entire PACS wavelength range.
Both lines were detected in emission (Fig.~\ref{f:ciioi}), but are unresolved at the 86\,km\,s$^{-1}$ (at 63\,$\mu$m) and 239\,km\,s$^{-1}$ (at 158\,$\mu$m) per resolution channel of PACS. Since the field is spatially resolved into $5\times5$ spaxels (`spectral pixels', each of side 9.4\arcsec), we can clearly see that the emission is centred on the star, and not due to an offset background object (Fig.~\ref{f:3x3}). 

\begin{figure}
   \centering
   \includegraphics[width=9.0cm]{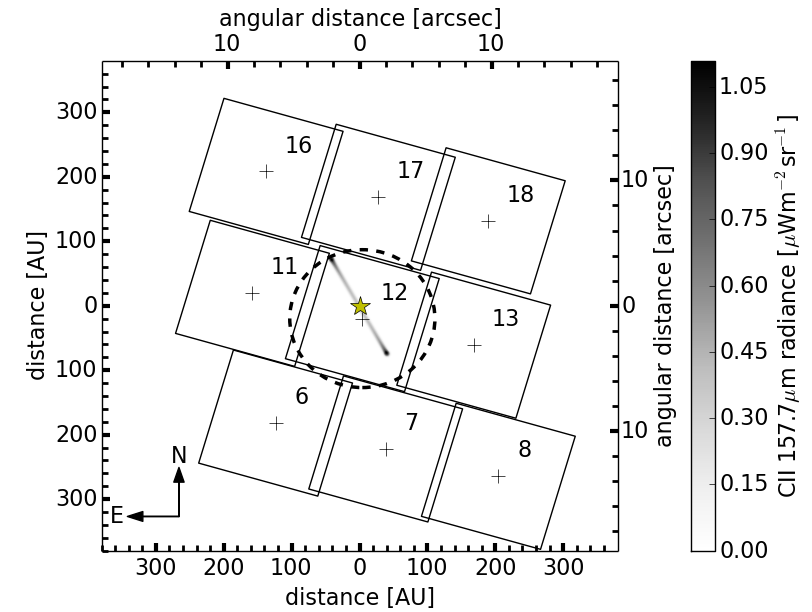}
      \caption{The central 3$\times$3 spaxels of the 5$\times$5 PACS array, showing the coverage of a modelled disk (model `5$\times$5\,AU') and location of the star with respect to the spaxel coordinates. The central position of each spaxel is marked by a +, and the position of $\beta$\,Pic at the epoch of the observation with a star. The dashed circle corresponds
to the 11\arcsec\ FWHM of the beam at 158\,$\mu$m.}
   \label{f:spaxels}
   \end{figure}

Fig.~\ref{f:spaxels} shows the orientation of the spaxels with respect to $\beta$\,Pic in the case of the \ion{C}{ii} observations. The telescope pointing model indicates that the central spaxel was offset from  $\beta$\,Pic by \upd{1.0}\arcsec\ for the \ion{C}{ii} and \upd{0.8}\arcsec\  for the \ion{O}{i} observations. The typical 68\,\%-confidence pointing accuracy for \textit{Herschel} at these epochs were $\sim$2\arcsec\ \citep{san14}.

The data reduction was essentially done as described in \citet{cat15}. We used the `background normalisation' pipeline script within the \textit{Herschel} interactive processing environment (HIPE) version 14.0 \citep{ott10}. This pipeline uses the background emission from the telescope itself to background subtract and calibrate the data. The data were binned into a wavelength grid by setting \texttt{oversample\,=\,4} (increasing the spectral sampling) and \texttt{upsample\,=\,1} (keeping neighbouring data points uncorrelated). The continuum was subtracted by fitting linear polynomials to the spectra with the line region masked. We used the pipeline-generated noise estimate `stddev' as relative weights for the continuum fit. This is particularly important for the \ion{O}{i} spectra, where the noise is smallest at the line centre and increases towards the spectral edges. Finally, we estimated the noise for each spaxel individually. To this end, we computed the standard deviation within two spectral windows placed sufficiently far from the line centre to avoid any contamination by line emission. The width of each window was chosen equal to 2.5$\times$ the spectral resolution of PACS (i.e.\ 2.5$\times$ the full width at half maximum (FWHM) of an unresolved line).

In order to measure the line flux, we fit a Gaussian function to the spectra, where the FWHM of the Gaussian was fixed to the FWHM of the line-spread function (i.e., the spectral resolution). The central wavelength of the Gaussian was also generally fixed to the expected wavelength corrected for the known radial velocity of $\beta$\,Pic \citep{bra11}. An exception was the central spaxel of the \ion{C}{ii} observations, where it is apparent by eye that there is a shift between the center of the emission line and the expected central wavelength (Fig.~\ref{f:ciioi}). In this case we let the line centre be a free parameter, and found the line centre
to be redshifted by 0.02\,$\mu$m. This shift is within specification of the PACS wavelength calibration, and would be expected for an off-centre point source (PACS observer's manual v.2.5.1, HERSCHEL-HSC-DOC-0832, \S\,4.7.2).

In order to estimate the error on the measured line flux, we produced \upd{$10^5$} fits\upd{, including the continuum fit and subtraction,} to resampled data for each spaxel with detected line emission. The resampling is achieved by adding normally distributed noise with a standard deviation according to the previously determined error of the spectrum, with the line centre and FWHM fixed. The error on the flux is then estimated as the standard deviation of the resulting distribution of fitted fluxes.

\section{Results and analysis}
\subsection{Detected emission\label{s:results}}

The [\ion{C}{ii}] emission detected by \textit{Herschel}/PACS from $\beta$\,Pic (Table~\ref{t:flux}) is $\sim$30\,\% lower than the corresponding emission detected by \textit{Herschel}/HIFI \citep[$(2.4\pm0.1)\times10^{-14}$\,erg\,s$^{-1}$cm$^{-2}$beam$^{-1}$,][]{cat14}, and a factor of three below the tentative detection by the \textit{Infrared Space Observatory} \citep{kam03}.  The difference in the observed flux between PACS and HIFI may be due to slight pointing differences, as the source likely is marginally resolved by a 11\arcsec\ beam; the precise pointing could thus be important. However, we were not able to reproduce this 30\,\% difference by varying the pointing centre by a few arcseconds (the expected pointing accuracy) in synthetic observations of our models, only a few \%. This could be because our assumed angular flux distribution is  wrong (e.g., the real distribution is not symmetric between the north-east [NE] and south-west [SW]), and/or because the pointing offset between the observations is higher than assumed. It could also be due to
an absolute calibration issue, which is expected to be 12\,\% but indeed could be off by up to 30\,\% (PACS observer's manual v.2.5.1, \S\,4.10.2.1). Finally, the difference could also be due to the beams of
PACS and HIFI not being identical.

\begin{table*}
\caption{\label{t:model}Model properties}
\centering
\begin{tabular}{lccccc}
\hline\hline
Model & $>5$\,AU\tablefootmark{a} & $>35$\,AU\tablefootmark{a} & 1$\times$1\,AU\tablefootmark{b} & 5$\times$5\,AU\tablefootmark{b} & 10$\times$10\,AU\tablefootmark{b} \\
\hline
$n_{\mathrm{peak,O}}$~[cm$^{-3}$]\tablefootmark{c} & $2.7\times10^5$ & $2.9\times10^4$ & $5.8\times10^6$ & $1.9\times10^4$ & $2.6\times10^4$ \\
$n_{\mathrm{peak,C}}$~[cm$^{-3}$]\tablefootmark{c} & $1.9\times10^3$ & $520$ & $1.7\times10^7$ & $3.7\times10^4$ & $1.1\times10^4$ \\
$n_{\mathrm{peak,e}}$~[cm$^{-3}$]\tablefootmark{c} & $1700$ & $320$ & $9.2\times10^4$ & $2200$  & $1100$ \\
$T_{\mathrm{peak}}$~[K]\tablefootmark{c} & 54 & 50 & 870 & 104 & 76 \\
C/O\tablefootmark{d} & 0.007 & 0.018 & 3.0 & 1.9 & 0.43 \\
O mass $[M_{\oplus}]$ & 1.052 & 0.415 & 0.052 & 0.004 & 0.023 \\ 
C mass $[M_{\oplus}]$ & 0.005 & 0.006 & 0.118 & 0.006 & 0.008 \\ 
$F_{\mathrm{obs,O}}/F_{\mathrm{mod,O}}$\tablefootmark{e} & 51 & 43 & 1.0 & 2.5 & 6.5 \\
$N_{\ion{O}{i}}~[$cm$^{-2}]$\tablefootmark{f} & $1.0 \times 10^{20}$ &2.1 $\times 10^{19}$ & $9.2 \times 10^{19}$ & $1.6 \times 10^{18}$ & $4.1 \times 10^{18}$ \\
$N_{\ion{C}{i}}~[$cm$^{-2}]$\tablefootmark{f} & $2.3 \times 10^{17}$ & $1.8 \times 10^{17}$ & $2.8 \times 10^{20}$ & $2.7 \times 10^{18}$ & $1.6 \times 10^{18}$\\
$N_{\ion{C}{ii}}~[$cm$^{-2}]$\tablefootmark{f} & $4.7 \times 10^{17}$ & $2.0 \times 10^{17}$ & $1.5 \times 10^{18}$ & $2.3 \times 10^{17}$ & $2.3 \times 10^{17}$\\
\hline
\end{tabular}
\tablefoot{
\tablefoottext{a}{Models with a spatial distribution according to \citet{nil12}; see Sect.\,\ref{s:model}.}
\tablefoottext{b}{Models with a torus-like spatial distribution; see Sect.\,\ref{s:model}.}
\tablefoottext{c}{The quantity measured at peak gas density; for the \citet{nil12} profiles this corresponds to the inner edge, for the tori models
the maximum is located in the midplane at 85\,AU.}
\tablefoottext{d}{The solar C/O is 0.5, and for carbonaceous meteoroids 0.1 \citep{lod03}.}
\tablefoottext{e}{The ratio between the observed and modelled flux in [\ion{O}{i}] 63\,$\mu$m.}
\tablefoottext{f}{The column density in the mid-plane against the star.}
}
\end{table*}

Oxygen has not previously been observed in emission, but a model of the $\beta$\,Pic gas disk by \citetalias{zag10} predicted a flux 3 orders of magnitude below the detected level. They assumed a spatial distribution of well-mixed gas derived from observations of \ion{Na}{i} \citep{bra04}, with only C overabundant \citepalias[by a factor of 20,][]{rob06}. \citet{cat14} used the HIFI observations of 
[\ion{C}{ii}] 158\,$\mu$m to update the \citetalias{zag10} model, and found that a C overabundance by a factor of 300 was required to explain the strong emission observed, assuming a well-mixed gas. To test how the observed  [\ion{O}{i}] 63\,$\mu$m emission constrains the mass and distribution of O, we produced a range of models assuming different spatial distributions and abundances of the gas, as reported in Sect.\,\ref{s:model} below.

\subsection{Gas disk models\label{s:model}}

To model the gas emission from the $\beta$\,Pic disk we used the \textsc{ontario} code, specifically developed for modeling gas in debris disks around A--F stars \citepalias{zag10}. Given the input parameters, which are the stellar spectrum and the spatial distribution of dust and gas (including its abundance), \textsc{ontario} computes the thermal and ionisation balance of the gas, and then the statistical equilibrium population of energy levels in species of interest, including their emitted spectra. This is in particular important for O that, in contrast to C, has level populations that generally are out of local thermal equilibrium (LTE) in the low electron density environment of the disk gas (\upd{we find the critical electron densities at 100\,K for \ion{C}{ii} and \ion{O}{i} to be 6\,cm$^{-3}$ and 2.5$\times 10^5$\,cm$^{-3}$, respectively}). Moreover, the [\ion{O}{i}] 63\,$\mu$m line, like [\ion{C}{ii}] 158\,$\mu$m, becomes optically thick at column densities of $\sim5\times10^{17}$\,cm$^{-2}$ (assuming a line width of 2\,km\,s$^{-1}$). The line luminosities of the disk will thus not only depend on the gas mass, but also strongly on the spatial distribution of the gas. 
 Since the 
[\ion{C}{ii}] 158\,$\mu$m and [\ion{O}{i}] 63\,$\mu$m lines are also important cooling lines, we updated the thermal and statistical 
equilibrium solution in \textsc{ontario} to approximate the radiative transfer effect using photon escape probabilities, essentially following Appendix~B of \citet{tie85}. We start by computing the level populations for the innermost grid points, and then continue outwards while adjusting for the extinction from interior grid points towards the star. As a second-order correction, we take into account locally scattered photons by using a photon escape formulation. The photon escape fraction is computed iteratively by starting with an assumption of complete escape, and then using the computed level population from the previous iteration to approximate the photon escape probability from the average of four directions  \citep[in, out, up, and down, as described in][]{gor04}, obtaining a new level population. Once the level populations of each grid point are defined, the radiative transfer equation is integrated along the line of sight from the observer, assuming a Keplerian velocity field to produce an angular distribution of the spectral profile, as described in \citet{cat14}.

\subsubsection{The assumed spatial distribution of gas\label{s:spatial}}

If the C and O are produced principally from the enhanced collision rate responsible for the production of CO observed by ALMA at an orbital radius of $\sim$85\,AU \citep{den14}, then one would expect the gas distribution to peak at 85\,AU and diffuse to other regions of the disk. Given that the CO molecule is short lived ($\sim$120\,yr at 85\,AU) while C and O in their ion/atomic form are stable, we would expect C and O to be more evenly distributed than CO, perhaps in a ring slowly diffusing away from the production orbital radius. If, on the other hand, the C and O are primarily produced by the same mechanism that is responsible for the observed metallic species in the disk \citep[Na, Fe, Ca, etc.,][]{bra04}, then one could expect the C and O spatial distribution to more closely follow the gas distribution inferred from the spatially resolved \ion{Fe}{i} emission \citep{nil12}, with the density
\begin{equation}
\label{eq:Fe_distr}
n(r,z) = n_0\left[\frac{2}{(r/r_0)^{2\alpha}+(r/r_0)^{2\beta}}\right]^{1/2} \exp\left[-\left(\frac{z}{h(r)}\right)^{\gamma}\right],
\end{equation}
where $h(r) = h_0(r/r_0)^{\delta}$ and the parameters $n_0$, $h_0$, $r_0$, $\alpha$, $\beta$, $\gamma$, and $\delta$ as listed in Table~3 of \citet{nil12}. One important parameter missing from Eq.~\ref{eq:Fe_distr} is the inner truncation radius of the disk, since the density $n(r,z)$ diverges as $r\to0$. Unfortunately, this parameter is difficult to assess due to scattered light residuals from the nearby star, dominating the noise inside 2\arcsec (40\,AU) in the \citet{nil12} observations. With VLT/UVES, \citet{bra04} were able to trace the \ion{Na}{i} and \ion{Fe}{i} a bit further in, to the limit of the observations at 0.7\arcsec (13\,AU) from the star. They found an asymmetry in that the disk on the NE side appears to rise in brightness all the way in, while the SW side shows a significant decrease in density inside 36\,AU.

The  \citet{cat14} \textit{Herschel}/HIFI \textit{spectrally} resolved observations of the [\ion{C}{ii}] 158\,$\mu$m emission line indirectly put constraints on the C \textit{spatial} distribution under the assumption of a Keplerian rotating disk. They found that the observations were best fit if the gas distribution inferred from \ion{Fe}{i} had an inner truncation radius between 30 and 100\,AU. Indeed, a single torus of radius $\sim$100\,AU could by itself reasonably fit the data \citep[see Fig.~4 in][]{cat14}.

Since the constraint on the mass of C and O in the disk depends on the assumed spatial distributions of the gas, we explore a few different reasonable configurations: 
\begin{enumerate}
\item A C/O gas that is well mixed with the observed Fe and Na distribution, using Eq.~\ref{eq:Fe_distr} with parameters fit from \citet{nil12}, C/O abundances as free parameters and the inner truncation radius at 5\,AU and 35\,AU (labelling the models `$>$5\,AU' and `$>$35\,AU').
\item Assuming C and O are produced from CO independently from Fe/Na, we use a torus of radius 85\,AU with a selection of assumed widths and scale heights,
motivated by the ALMA CO observations. The model tori have scale heights that are equivalent to their radial extents, with the full-width half-maxima being 1\,AU, 5\,AU and
10\,AU according to a Gaussian distribution with the peak density of C and O being free parameters. The models are labelled `1$\times$1\,AU', `5$\times$5\,AU', and `10$\times$10\,AU', respectively.
\end{enumerate}
The principal purpose of these models is not to accurately model the real spatial distribution of the gas, which indeed likely departs from the assumed cylindrical symmetry implied here. The
purpose is instead to study how the predicted emission is influenced by assumptions on the spatial distribution and abundance of gas, to help us distinguish between possible scenarios. To
make a realistic model of the C and O gas, their spatial distribution would have to be better constrained. Properties of the models are listed in Table~\ref{t:model}.

\subsubsection{Comparing models to observation}

To compare the models with data, we computed a grid of models, deriving the corresponding angular distribution of line luminosity in the sky \citep[as in][]{cat14}, and then used synthetic PACS
observations \citep[as described in][but with the updated v6 of the PACS spectrometer beams]{cat15} to produce a model spectrum for each spaxel to be compared with the observed data. Fig.~\ref{f:3x3} shows the central 3$\times$3 spaxels of one such model overplotted on the observed [\ion{C}{ii}] 158\,$\mu$m emission.

We found it challenging to reproduce the detected  [\ion{O}{i}] 63\,$\mu$m emission while at the same time being consistent with the observed
[\ion{C}{ii}] 158\,$\mu$m emission. Under the assumption of C and O being well mixed together, most spatial distributions tend to underproduce the 63\,$\mu$m line
in comparison to the 158\,$\mu$m emission. The complication arises from the complex interplay between the heating/cooling, excitation mechanisms and optical thickness effects. Increasing the O abundance increases the cooling, but only until the dominant 63\,$\mu$m cooling line becomes optically thick. Increasing the abundance even further may then \textit{decrease} the energy output in the 63\,$\mu$m line, as locations with lower electron density become optically thick and block the radiation. We thus find that there is a C/O ratio that will maximise the 63\,$\mu$m flux for a fixed 158\,$\mu$m flux, and for a range of models we tested, this maximum 63\,$\mu$m flux was below the observed flux. For the 5 different spatial distributions we considered (listed in Table~\ref{t:model}), we varied the C and O abundances freely, but filtered out any model that gave a
synthetic [\ion{C}{ii}] 158\,$\mu$m flux that differed more than 5\,\% from the observations. We then picked the model that comes closest to reproduce the [\ion{O}{i}] 63\,$\mu$m emission. Only the densest 1$\times$1\,AU model is able to fully reproduce the 63\,$\mu$m line, while other models underpredict the emission by a factor 2.5--51 (Table~\ref{t:model}).

Another problem with the models presented in Table~\ref{t:model} is that the implied column density against the star is orders of magnitude larger than the column densities observed by
FUSE \citepalias{rob06}, which are $N_{\ion{O}{i}} = (3-8)\times10^{15}$\,cm$^{-2}$, $N_{\ion{C}{i}} = (2-4) \times 10^{16}$\,cm$^{-2}$, and $N_{\ion{C}{ii}} = (1.6-4.1) \times 10^{16}$\,cm$^{-2}$.
As discussed by \citet{bra11}, part of the difference could be due to the difficulties in measuring unresolved optically thick lines. In \citetalias{rob06}, a single broad component was assumed for an \ion{O}{i} absorption line (broadening parameter $b = 15$\,km\,s$^{-1}$), which effectively gives a lower limit on the column density. If instead a narrow component is assumed, say $b = 1-2$\,km\,s$^{-1}$, the column density could be much higher (in excess of $N_{\ion{O}{i}} = 10^{17}$\,cm$^{-2}$) and still be consistent with the observed line profile \citep[see supplementary Fig.~1 of][]{rob06}. The lower \ion{C}{i} column density is harder to explain in this way, since it is based on a robust measurement of the $^3P_0$ state, where an optically thin line was observed by STIS \citep{rob00}. We conclude that the absorption measurements are consistent with a low C/O ratio gas ($\lesssim1$) in a non-cylindrically symmetric distribution.

\section{Discussion}

From the modeling described in the previous section we conclude that we can only explain the observed 63\,$\mu$m emission if there is a region in the gas disk that is sufficiently dense to effectively excite O, meaning an electron density \updB{$n_{\mathrm{e}} \gtrsim 2.5\times10^4$}\,cm$^{-3}$, the critical density for the [\ion{O}{i}] 63\,$\mu$m line. This region does not have to be very large, however; if optically thick, in LTE, and
with temperatures in the range 80--100\,K, a high-density clump with a diameter of a few AU would be sufficient to produce the observed emission. Such a clump would show up in the
angular distribution of [\ion{C}{ii}] 158\,$\mu$m and in particular in [\ion{C}{i}]. Considering the clump recently imaged in CO by ALMA \citep{den14}, perhaps there is a corresponding clump in the
C and O distribution that would explain the enhanced [\ion{O}{i}] 63\,$\mu$m emission. The [\ion{C}{ii}] 158\,$\mu$m spectral profile by HIFI also seems consistent with a C clump to the SW \citep[see Fig.~6 of][]{cat14}. 
A clumpy distribution may thus be preferred by observations, but it is in contrast to the expectation outlined in Sect.\,\ref{s:spatial}, that C and O would have diffused into a smooth distribution due to their much longer lifetime than CO. This could be resolved if the diffusion process is slower than anticipated
\citep[$\alpha\lesssim0.01$, resulting in a viscous timescale $\gtrsim$ Myr,][]{xie13} so that the pattern imprinted by the CO distribution lasts longer, or if the event producing CO is recent ($\ll$ Myr). More detailed models are required to evaluate if these scenarios are credible.

A region of enhanced electron density to the SW might be able to explain another puzzling property of the $\beta$\,Pic disk; why the asymmetry between the NE and SW is so pronounced in \ion{Na}{i} and \ion{Fe}{i} \citep[see Figs.~2 \& 3 of][]{bra04}. With an increased electron density, the neutral fractions of these would dramatically rise due to more frequent recombination,
meaning that any given particle would spend a longer time in its neutral state. Since the radiation force on the neutral species of Fe and Na is much stronger than gravity (27$\times$ and 360$\times$, respectively) and only the ionised species are effectively braked by Coulomb interaction with the C$^+$ \citep{fer06}, the atoms will be removed from the system with a drift velocity 
$\varv_{\mathrm{drift}} = \varv_{\mathrm{ion}}f$, where $\varv_{\mathrm{ion}}$ is the average velocity the atom reaches before ionisation and subsequent braking by the C$^+$ gas, and $f$ is the fraction
of the time the particle spends in its neutral state. For Fe,  $\varv_{\mathrm{ion}} = 0.5$\,km\,s$^{-1}$ and for Na,  $\varv_{\mathrm{ion}} = 3.3$\,km\,s$^{-1}$ \citep{bra11}.

The spatial distribution of C and O is presently not constrained enough to be able to derive an accurate O mass. If the C and O are indeed produced by the outgassing and subsequent dissociation 
of molecules from colliding comets, as suggested by \citet{zuc12} and \citet{den14}, then the expected C/O ratio would be 0.1--1, depending on the fraction of CO/H$_2$O present in the comet-like bodies.
Upcoming ALMA observations of the [\ion{C}{i}] 609\,$\mu$m line should be able to settle the case. 

\section{Concluding summary}
In summary, our conclusions are as follows:
   \begin{enumerate}
      \item Emission from \ion{C}{ii} and \ion{O}{i} has been detected from $\beta$\,Pic.
      \item The detected emission from \ion{O}{i} 63\,$\mu$m is much stronger than expected from cylindrically symmetric models.
      \item A region of relatively high density, perhaps in a clump similar to the one observed in CO, is required to explain the \ion{O}{i} 63\,$\mu$m emission.
	\item To derive a reliable mass of O, and thereby constrain the C/O ratio of the disk, knowing the spatial distribution of C and O is essential.
   \end{enumerate}

\begin{acknowledgements}
We thank Aki Roberge, Philippe Th\'ebault and Yanqin Wu for helpful discussions. AB was supported by the Swedish National Space Board (contract 75/13).
BA was a Postdoctoral Fellow of the Fund for Scientific Research, Flanders.
RJI acknowledges support from ERC in the form of the Advanced Investigator Programme, 321302, COSMICISM.
PACS has been developed by a consortium of institutes led by MPE (Germany) and including UVIE (Austria); KU Leuven, CSL, IMEC (Belgium); CEA, LAM (France); MPIA (Germany); INAF-IFSI/OAA/OAP/OAT, LENS, SISSA (Italy); IAC (Spain). This development has been supported by the funding agencies BMVIT (Austria), ESA-PRODEX (Belgium), CEA/CNES (France), DLR (Germany), ASI/INAF (Italy), and CICYT/MCYT (Spain).
\end{acknowledgements}

\bibliographystyle{aa}  % style aa.bst 
\bibliography{bPicPACS16} 

\end{document}